\documentclass[letterpaper]{article}

\usepackage{aaai2026}
\nocopyright

\usepackage{times}
\usepackage{helvet}
\usepackage{courier}
\usepackage[hyphens]{url}
\usepackage{graphicx}
\usepackage{natbib}
\usepackage{caption}
\usepackage{booktabs}
\usepackage{longtable}
\usepackage{array}
\usepackage{pdflscape}
\usepackage{ragged2e}

\usepackage{algorithm}
\usepackage{algorithmic}

\usepackage{newfloat}
\usepackage{listings}
\DeclareCaptionStyle{ruled}{labelfont=normalfont,labelsep=colon,strut=off}
\floatstyle{ruled}
\newfloat{listing}{tb}{lst}{}
\floatname{listing}{Listing}

\title{Whose Good, Whose Place?\\
The Moral Geography of Agentic AI for Social Good}

\author{
Poli Nemkova\textsuperscript{\rm 1},
Haeshitha Indukuri\textsuperscript{\rm 1},
Jaedon Charles\textsuperscript{\rm 2}
}

\affiliations{
\textsuperscript{\rm 1}University of North Texas\\
\textsuperscript{\rm 2}Florida International University\\
\texttt{poli.nemkova@unt.edu},
\texttt{HaeshithaIndukuri@my.unt.edu},
\texttt{jchar144@fiu.edu}
}

\begin{document}

\maketitle

\begin{abstract}
Agentic AI systems are increasingly proposed for social-good domains,
invoking the United Nations Sustainable Development Goals (SDGs) as a
vocabulary of global benefit. Yet invocation of social good does not
establish accountability to the communities a system claims to serve.
We present a structured survey of 112 papers on agentic AI for social
good published between 2015 and 2026, and report a \textit{moral-geographic
asymmetry}: the field's contextual underspecification is not uniform---it
is concentrated precisely where contextual stakes are highest.

Each paper is coded across domain, SDG alignment, agent architecture,
autonomy level, agent goal, evaluation type, deployment status, and
geographic focus, using a dual-LLM annotation procedure (GPT-4o and
Llama-3.3-70B) with stratified human validation on 50 papers.
Across the corpus, 82 of 112 papers (73\%) specify no geographic context.
This absence is sharply uneven: papers aligned with health or
physical/ecological SDGs specify a geography 37--40\% of the time,
while papers aligned with institutional and social-policy SDGs do so
only 13\%---and 0\% under strict coding that counts only papers naming
a concrete country, city, region, or population
($\chi^2(2) = 8.78$, $p = .012$, Cram\'{e}r's $V = .283$;
strict: $\chi^2(2) = 7.60$, $p = .022$, $V = .263$).
The pattern holds across seven out of  eight sensitivity configurations tested,
including alternative SDG cluster assignments and single-SDG subsets.
SDG~16---peace, justice, and strong institutions---is both the
most-covered goal in the corpus and the one with the lowest
geographic-specification rate (13\% broad; 2.6\% strict).

We interpret this as moral abstraction: agentic AI for social good
treats institutional good as universal in ways it does not treat health
or ecological good, precisely where local political, legal, and cultural
context matters most. A second finding compounds this: only 28 of 112
papers (25\%) report any real-world deployment or small-scale test,
and this gap is uniform across architecture, era, and geographic
specificity---a field-wide condition, not a localized one.
We identify five accountability gaps and propose a minimal reporting
standard for more context-specific, participatory, and accountable
agentic AI for social good.
\end{abstract}

\section{Introduction}
Agentic AI has become one of the dominant imaginaries of contemporary
artificial intelligence. Large language models are increasingly embedded in
systems that plan, coordinate, monitor, retrieve evidence, allocate
resources, persuade users, or act through tools. In parallel, researchers
have begun to propose agentic AI systems for social-good domains including
disaster response, public health, education, social policy, sustainability,
online safety, urban planning, agriculture, and humanitarian
decision-making.

This emerging literature inherits the moral vocabulary of AI for social
good, often invoking the United Nations Sustainable Development Goals
(SDGs) as a framework for global benefit. Yet the invocation of social good
does not by itself establish accountability to the communities a system
claims to benefit. A system can target an SDG while failing to name the
communities, institutions, or geographic settings in which it would operate.
It can describe agents for planning, coordination, monitoring, allocation,
or persuasion while leaving open who sets the goals, who evaluates the
outcomes, and who bears the risks of failure.

This paper asks: \emph{whose good is agentic AI for social good being built
to serve?} Rather than treating social good as a self-evident label, we
examine how this literature allocates attention across domains, SDGs,
geographies, architectures, autonomy levels, evaluation methods, deployment
settings, and agent goals. We use the phrase \emph{moral geography} to
refer to this distribution of attention: the problems, places, and
populations that become legible as sites for agentic AI intervention, and
those that remain peripheral or unnamed.

We present a structured survey of 112 papers published between 2015 and
2026, annotated by a dual-LLM coding procedure (GPT-4o and Llama-3.3-70B)
with human validation on a stratified 50-paper subsample. Each paper is
coded along eight dimensions: domain, SDG alignment, agent type, autonomy
level, real-world deployment status, evaluation type, agent goal, and
geographic focus. This design allows us to move beyond a narrative review
and empirically examine how the field frames its social-good claims.

Our central finding is a moral-geographic asymmetry in contextual
specification. Across the corpus, 82 of 112 papers (73\%) do not specify any
geographic context for the proposed system. This absence is not evenly
distributed across the SDGs: papers aligned with health or
physical/ecological SDGs specify a geography 37--40\% of the time, while
papers aligned with institutional and social-policy SDGs do so only 12.8\%
of the time ($\chi^2(2) = 8.78$, $p = 0.012$, Cramér's $V = 0.283$). The
pattern is sharpest for SDG~16, peace, justice, and strong institutions:
the most-covered goal in the corpus and the one with the lowest
geographic-specification rate.

This asymmetry matters because institutional interventions are especially
context-dependent. Systems that plan, allocate, monitor, persuade, or
coordinate within governance settings depend on local political, legal,
cultural, and institutional conditions. We therefore read the pattern as a
form of moral abstraction: agentic AI for social good treats institutional
good as universal in a way that it does not treat health or ecological
good.

A second finding concerns the gap between proposal and practice. Only 28 of
112 papers (25\%) report any form of real-world deployment or small-scale
test; the remaining 75\% are conceptual or simulation-only. This gap is not
concentrated in one subset of the corpus: papers that name a geography do
not deploy at higher rates, post-2023 LLM-agent papers do not deploy at
lower rates than earlier multi-agent-systems work, and LLM-based
architectures do not deploy at lower rates than classical agent
architectures. On the evidence available, the proposal-to-practice gap is a
field-wide condition.

We make three contributions. First, we provide a structured map of agentic
AI for social good across domains, SDGs, architectures, autonomy levels,
agent goals, evaluation types, deployment status, and geographic focus.
Second, we identify a moral-geographic asymmetry: contextual abstraction is
concentrated in the institutional and social-policy SDGs that the field
most often invokes. Third, we propose an accountability agenda for agentic
AI for social good, arguing that future work should specify target
contexts, justify architectural complexity, involve affected communities,
document failure modes, and align evaluation practices with the strength of
its social claims.

\section{Background: From AI for Social Good to Agentic AI for Social Good}

\label{sec:related}

\subsection{Mapping AI for Social Good}

A substantial literature has mapped how AI techniques are being applied to
social good domains. Early surveys provided broad taxonomies of AI4SG
applications across health, environment, education, poverty, crisis response,
and governance \citep{shi2020ai, tomasev2020ai}. \citet{floridi2020design}
proposed seven essential factors for AI4SG design, and
\citet{cowls2021good} examined how AI projects align with the United
Nations Sustainable Development Goals (SDGs). \citet{vinuesa2020role}
systematically assessed how AI could enable or inhibit each of the 17 SDGs
and 169 targets, while \citet{saetra2022ai} examined the indirect and
second-order effects of AI on the SDG agenda. More recent work has surveyed
NLP for social good specifically \citep{karamolegkou2024nlp4sg}.

These prior surveys share two features that bound their contributions and
motivate ours. First, they treat AI as a broad family of techniques without
distinguishing systematically among predictive, generative, and agentic
systems. Second, they ask whether AI \emph{could} contribute to SDGs, or what
factors make AI4SG projects succeed, rather than empirically auditing how
specific subfields of AI \emph{describe} their target communities, contexts,
and accountability structures. Our survey addresses the second gap for the
subfield of agentic AI: we ask not whether agentic AI could serve social
good in principle, but how the literature already proposing it for social
good situates its systems in concrete settings.

\subsection{Agentic AI and the Stakes of Agency}

Traditional AI4SG systems often centered on prediction, classification, or
optimization. Agentic systems are increasingly described as systems that
plan, coordinate, communicate, monitor, retrieve information, use tools, or
act over time. Multi-agent systems have a longer history in domains such as
traffic, disaster response, public safety, agriculture, and resource
allocation \citep{wooldridge2009introduction, stone2000multiagent,
ramchurn2012disaster, tambe2011security}. More recently, LLM agents and
multi-agent LLM systems have introduced a new wave of general-purpose
agentic proposals framed around reasoning, delegation, tool use, and
human--AI collaboration \citep{yao2023react, park2023generative, xi2023rise,
wang2024survey}. Recent surveys of agentic AI focus on architectures and
capabilities \citep{xi2023rise, wang2024survey}, value alignment
\citep{zeng2025multilevel}, and general-purpose applications. None, to our
knowledge, systematically audits the social-good claims made by the
agentic-AI literature.

The shift from predictive to agentic systems matters ethically. The more a
system is framed as an agent, the more important it becomes to ask what
goals it pursues, whose preferences it represents, what authority it has
to act, and how its actions can be contested or corrected
\citep{floridi2018ai4people, dobbe2021system}. An advisory agent for
clinical workflow raises different accountability questions than a fully
autonomous coordination system for disaster response. A persuasive agent
for civic engagement raises different concerns than a verification agent
for misinformation. These distinctions are not exhausted by general
critiques of AI4SG; they require attention to architecture, autonomy, and
deployment in their own right.




\section{Method}
\subsection{Corpus Construction}

We constructed a corpus of agentic-AI-for-social-good papers published
between 2015 and 2026, with 91\% appearing from 2020 onward. The corpus spans
multi-agent systems, human-agent collectives, simulation agents, embodied
agents, single-agent systems, LLM agents, multi-agent LLMs, and recent
agentic-AI workflows.

We followed a PRISMA-style systematic review procedure~\cite{page2021prisma},
searching four academic discovery platforms---Google Scholar, ACL Anthology,
Semantic Scholar, and Elicit---with queries organized around all 17 United
Nations Sustainable Development Goals to reduce retrieval bias. For each SDG
we used both application-oriented queries (e.g., \emph{clinical triage agents},
\emph{disaster response agents}) and task-oriented queries (e.g.,
\emph{misinformation verification}, \emph{resource allocation},
\emph{monitoring}). Retrieved records were merged and deduplicated,
yielding 177 candidates.

Papers were included if they: (1) described, analyzed, evaluated, or
surveyed an AI agent, multi-agent system, LLM agent, or human--AI
collaboration; (2) targeted a domain plausibly connected to social good,
sustainable development, or one or more SDGs; and (3) provided enough
information to code domain, agent type, and social-good orientation. We
excluded policy-only papers with no agentic system, non-agentic ML systems,
commercially motivated systems without a clear social-good objective, purely
theoretical work without an identifiable system, and papers whose title and
abstract were too generic to establish relevance. After dual-LLM screening
and human eligibility review, 65 papers were excluded, yielding a final
corpus of 112 papers.

We use \emph{agentic AI for social good} broadly to capture systems that
perform or support goal-directed behavior in socially consequential domains,
including systems that plan, coordinate, monitor, verify, persuade, allocate
resources, forecast, or support human decision-making.

\subsection{Annotation Scheme}

Each paper was annotated along eight dimensions: domain, SDG alignment, agent
type, autonomy level, deployment status, evaluation type, agent goal, and
geographic focus. Table~\ref{tab:annotation-schema} summarizes the scheme.

\begin{table}[t]
\centering
\small
\begin{tabular}{p{0.3\linewidth}p{0.6\linewidth}}
\toprule
\textbf{Field} & \textbf{Description} \\
\midrule
\texttt{Domain} & Primary social-good area addressed. \\
\texttt{UN\_SDGs} & SDG(s) most closely aligned with the paper (numbers
1--17, or \emph{N/A}). \\
\texttt{Agent\_Type} & Architectural category: multi-agent systems, LLM
agents, multi-agent LLMs, agentic AI systems, simulation agents, human--AI
teams, single-agent systems, embodied agents, or non-agent. \\
\texttt{Autonomy\_Level} & Degree of agency: advisory, decision-support,
mixed, fully autonomous, or not applicable. \\
\texttt{Real\_World\_ Deployment} & Implementation extent: conceptual,
simulation-only, small-scale test, field-deployed, or varied. \\
\texttt{Eval\_Type} & Primary evaluation mode: conceptual study, simulation,
offline benchmark, field study, user study, literature review, or mixed. \\
\texttt{Agent\_Goal} & Primary agent function: planning, coordination,
persuasion/engagement, resource allocation, verification, monitoring,
forecasting, mixed, or not applicable. \\
\texttt{Geographic\_ Focus} & Geographic context as stated in the paper.
Coded \emph{Not specified} unless a concrete geography or population is
explicitly named. For analysis, we distinguish three levels:
\emph{not specified} (no context given), \emph{global or broad} (e.g.,
``global,'' ``Sub-Saharan Africa''), and \emph{concrete} (named country,
city, region, or identified population). \\
\bottomrule
\end{tabular}
\caption{Annotation schema. The three-level \texttt{Geographic\_Focus}
distinction is used in sensitivity analyses; the binary
specified/not-specified coding replicates the main analysis.}
\label{tab:annotation-schema}
\end{table}

The scheme captures both what systems do technically and how concretely they
are situated socially. Deployment status, evaluation type, and geographic
focus in particular serve as indicators of how far a paper moves from an
abstract social-good claim toward an accountable sociotechnical intervention.

\subsection{Dual-LLM Annotation with Human Validation}
\label{sec:annotation}

Annotating 177 papers along eight dimensions by hand would have been
prohibitive, and a single LLM annotator would expose labels to one model's
idiosyncratic biases. We therefore adopted a \emph{dual-LLM annotation
procedure with stratified human validation} in three stages.

\paragraph{Stage 1: Dual-LLM annotation.}
Each candidate was independently labeled by \texttt{gpt-4o} (OpenAI) and
\texttt{llama-3.3-70b-versatile} (Groq API) using identical
controlled-vocabulary prompts (v2.0) at temperature 0.0. Responses were
validated against controlled vocabularies before being written to disk. The
prompt provided explicit definitions and decision heuristics for adjacent
categories (e.g., \emph{field-deployed} requires real end-users in their
operational context, distinct from \emph{small-scale test} for
recruited-participant studies). Both models labeled from title and abstract
only, without outside knowledge. Using two model families with the same
prompt isolates cross-model consistency from prompt design.

\paragraph{Stage 2: Cross-LLM agreement.}
We computed inter-annotator agreement on all 177 candidates
(Table~\ref{tab:iaa-llm}). Nominal categorical fields use Cohen's $\kappa$;
ordinal fields (\texttt{Autonomy\_Level}, \texttt{Real\_World\_Deployment})
use both unweighted and quadratic-weighted $\kappa$; the multi-label
\texttt{UN\_SDGs} field uses macro $\kappa$ over SDGs 1--16 and mean Jaccard
similarity; and open-vocabulary \texttt{Geographic\_Focus} uses exact-match
agreement after binarization. SDG~17 is excluded from macro $\kappa$ as it
lacks concrete subject matter and is structurally difficult to label from an
abstract.

\begin{table}[t]
\centering
\small
\begin{tabular}{lrr}
\toprule
\textbf{Field} & \textbf{Agreement} & \textbf{$\kappa$} \\
\midrule
\multicolumn{3}{l}{\textit{Nominal categorical (Cohen's $\kappa$)}} \\
\texttt{Agent\_Type}            & 83.1\% & 0.79 \\
\texttt{Eval\_Type}             & 82.5\% & 0.77 \\
\texttt{Agent\_Goal}            & 79.1\% & 0.74 \\
\midrule
\multicolumn{3}{l}{\textit{Ordinal (quadratic-weighted $\kappa$)}} \\
\texttt{Autonomy\_Level}        & 68.4\% & 0.73 \\
\texttt{Real\_World\_Deployment}$^{\dagger}$ & 80.1\% & 0.73 \\
\midrule
\multicolumn{3}{l}{\textit{Multi-label (macro $\kappa$ / Jaccard)}} \\
\texttt{UN\_SDGs} (macro $\kappa$, SDGs 1--16) & --- & 0.60 \\
\texttt{UN\_SDGs} (mean Jaccard) & --- & 0.66 \\
\midrule
\multicolumn{3}{l}{\textit{Open vocabulary}} \\
\texttt{Geographic\_Focus} (exact match) & 87.0\% & --- \\
\bottomrule
\end{tabular}
\caption{Cross-LLM agreement between GPT-4o and Llama-3.3-70B on the 177
candidates. $^{\dagger}$Weighted $\kappa$ on the 166 rows where both
annotators selected an on-scale label. All fields reach substantial agreement
(Landis and Koch, 1977); \texttt{UN\_SDGs} macro $\kappa$ is at the
substantial/moderate boundary.}
\label{tab:iaa-llm}
\end{table}

Agreement is substantial on all categorical fields and on binarized
\texttt{Geographic\_Focus} (87\%), the field carrying our main empirical
finding. The two annotators converge less strongly on \texttt{Autonomy\_Level}
(unweighted $\kappa = 0.56$, weighted 0.73); we treat this as a substantive
finding about field-level ambiguity rather than a methodological flaw, and
read autonomy patterns in section \textit{The Distance Between Proposal and Practice} descriptively.

\paragraph{Stage 3: Human validation.}
One author (blind to the LLM labels) coded a stratified 50-paper subsample
using the same controlled vocabularies, drawn from: (a) all 18 papers where
the LLMs disagreed on relevance; (b) 15 papers contesting the
high-definitional-stakes \texttt{Real\_World\_Deployment} field; and (c) 17
papers by stratified random sampling across \texttt{Agent\_Type} agreement
strata. Human--LLM agreement is reported in Table~\ref{tab:iaa-human}.

\begin{table}[t]
\centering
\small
\begin{tabular}{lrrrr}
\toprule
& \multicolumn{2}{c}{\textbf{GPT-4o}} &
  \multicolumn{2}{c}{\textbf{Llama-3.3-70B}} \\
\cmidrule(lr){2-3}\cmidrule(lr){4-5}
\textbf{Field} & \textbf{Agr.} & \textbf{$\kappa$} &
                 \textbf{Agr.} & \textbf{$\kappa$} \\
\midrule
\texttt{Agent\_Type}            & 72\% & 0.61 & 70\% & 0.59 \\
\texttt{Autonomy\_Level}        & 68\% & 0.51 & 74\% & 0.60 \\
\texttt{Real\_World\_Deployment}& 70\% & 0.48 & 70\% & 0.43 \\
\texttt{Eval\_Type}             & 74\% & 0.68 & 72\% & 0.66 \\
\texttt{Agent\_Goal}            & 72\% & 0.60 & 70\% & 0.58 \\
\texttt{Geographic\_Focus} (bin.)& 70\% & 0.44 & 68\% & 0.41 \\
\bottomrule
\end{tabular}
\caption{Human--LLM agreement on the stratified 50-paper subsample. Both
annotators show moderate-to-substantial agreement across all fields. GPT-4o
has the higher human-LLM $\kappa$ on five of six fields; Llama-3.3-70B is
higher on \texttt{Autonomy\_Level}.}
\label{tab:iaa-human}
\end{table}

\subsection{Consensus Label Construction}
\label{sec:consensus}

We combine the three label sources into a single consensus annotation per
field using: (1) the human label if the paper is in the 50-paper subsample;
(2) otherwise, the LLM with higher human-LLM $\kappa$---GPT-4o for all fields
except \texttt{Autonomy\_Level}, where Llama-3.3-70B scores higher.

Of the 177 candidates, 65 were excluded: 55 by both-LLMs consensus as
non-agent, 8 flagged during human eligibility review, and 2 pre-2015 papers.
The final 112-paper corpus derives 50 labels from the human annotator
(44.6\%) and 62 from the field-appropriate LLM (55.4\%).

One limitation of this procedure is that per-paper LLM disagreements outside
the human-validated subsample are not individually adjudicated. We chose this
over using a single LLM throughout because it preserves the
human-validation grounding; the cross-LLM agreement statistics in
Table~\ref{tab:iaa-llm} characterize the reliability of the LLM-only
portion. We note, additionally, that the human-LLM $\kappa$ values in
Table~\ref{tab:iaa-human} are estimated on a subsample deliberately enriched
for LLM disagreements, and therefore represent a conservative lower bound on
agreement over the full corpus.

\subsection{Quality Control and Epistemic Scope}
\label{sec:qc}

Because annotation relied on titles and abstracts rather than full papers,
our labels are evidence of how papers \emph{publicly frame} their
contributions, not exhaustive reconstructions of system details. This is
appropriate for a survey of field-level framing: the metadata most visible to
readers, funders, and downstream researchers is precisely what we measure.
We do not claim to know what every system did in practice; we claim to know
how it was presented.

This distinction matters most for geographic focus. A paper coded as
\emph{Not specified} may have been motivated by a concrete setting without
naming it in the abstract. Our claim is therefore about \emph{visible framing}:
the geography that appears in a paper's public face. This is still ethically
meaningful, because it is that framing---not the authors' unexpressed
intentions---that shapes how the field imagines and replicates work.

Conservative coding defaults were enforced throughout. When the abstract was
missing or uninformative, models were instructed to use \emph{Not specified},
\emph{Not applicable}, or \emph{No (conceptual)} rather than inferring
stronger claims. \texttt{Geographic\_Focus} required an explicitly named
geography; \texttt{Real\_World\_Deployment} required explicit evidence of
field deployment or real-user testing. After consensus construction, all 112
records were inspected for invalid labels, malformed SDG entries, and
duplicates. The binary \texttt{Geographic\_Focus} labels used in the main
analysis were additionally verified by hand across all 112 papers.

\subsection{Analytic Strategy}
\label{sec:analytic}

We analyze the corpus descriptively and inferentially. Descriptive analyses
report distributions across year, SDG, domain, agent type, autonomy, deployment,
evaluation, agent goal, and geographic focus. Inferential analyses report a
small number of pre-specified hypothesis tests. We use Fisher's exact test for
$2{\times}2$ tables and $\chi^2$ tests of independence for larger tables,
supplemented by Cramér's $V$ as an effect-size measure. Given the corpus size
($n=112$), we report effect sizes alongside $p$-values and do not apply
multiple-comparisons correction; all reported tests are pre-specified.

\paragraph{SDG cluster assignment.}
For the main moral-geographic asymmetry test, we group SDGs into three
clusters: \emph{health and welfare} (SDGs 1--3); \emph{physical and
ecological} (SDGs 6, 7, 11, 13--15); and \emph{institutional and social}
(SDGs 4, 5, 8--10, 12, 16, 17). When a paper carries multiple SDGs spanning
clusters, it is assigned to the cluster of its numerically lowest SDG. We
pre-specified two alternative cluster assignments as robustness checks:
Variant~A moves SDG~11 (sustainable cities) into the institutional cluster;
Variant~B moves SDG~8 (decent work) into the health/welfare cluster. We also
report results restricted to the 49 papers with unambiguous single-SDG
assignments.

\paragraph{Geographic coding.}
The primary analysis uses a binary coding: a paper is \emph{specified} if
any geographic context appears (including broad terms such as ``global'' or
``Sub-Saharan Africa''), and \emph{not specified} otherwise. This replicates
the most inclusive standard and matches the original coding instruction.
As a pre-specified sensitivity analysis we apply a \emph{strict} coding that
counts only papers naming a concrete country, city, region, or identified
population---explicitly excluding claims of global scope. ``Global'' claims
are treated as a third, distinct category in a three-way analysis: papers
claiming global applicability without naming a governance setting,
legal regime, or affected community are themselves an instance of the moral
abstraction the paper identifies. All three coding schemes are reported in
Table~\ref{tab:sensitivity}.

Normatively, we interpret distributions through the lens of accountability,
distributive attention, and contextual specificity, asking whether the field
provides adequate justification, context, and evidence for its social-good
claims---not whether equal attention across SDGs or regions would be the
correct allocation.

\section{The Moral Geography of Agentic AI for Social Good}
\label{sec:moral-geography}

\subsection{A Rapidly Expanding Post-2023 Field}

\begin{figure}[t]
    \centering
    \includegraphics[width=0.95\linewidth]{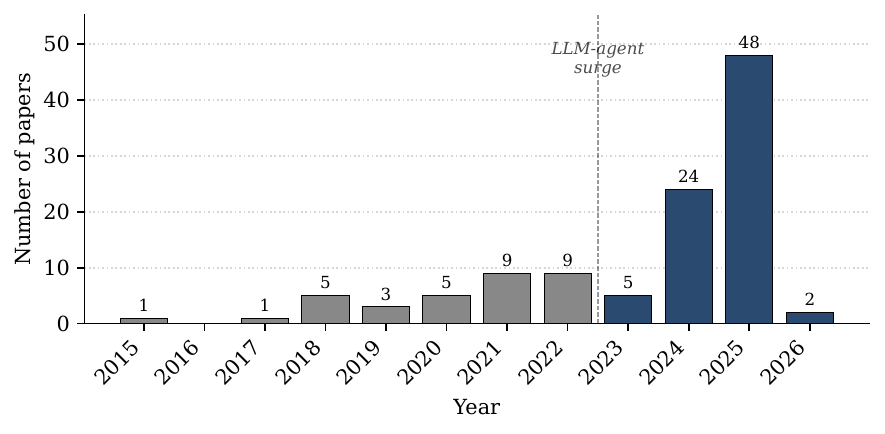}
    \caption{Distribution of corpus papers by publication year. The corpus
    is heavily concentrated in recent years, with most papers appearing
    after 2023.}
    \label{fig:papers-by-year}
\end{figure}

The corpus shows a sharp temporal concentration in recent years
(Figure~\ref{fig:papers-by-year}). Only 10 papers appear before 2020; the
field then accelerates substantially after the rise of LLM-based agentic
systems. Of the 112 papers, 79 (71\%) appear in 2023 or later and 74 (66\%)
in 2024--2026.

This is not simply an extension of older multi-agent systems research. Earlier
work focused on coordination, simulation, public safety, disaster response, and
resource allocation. Newer work uses the language of LLM agents, agentic
workflows, copilots, and multi-agent LLMs. The shift expands the apparent scope
of what agents can do while also raising the stakes of accountability:
general-purpose agentic systems can be proposed for many social-good domains
quickly, but speed of proposal does not guarantee contextual grounding,
community engagement, or real-world evaluation.

\subsection{Uneven SDG Attention}

\begin{figure*}[t]
    \centering
    \includegraphics[width=0.8\textwidth]{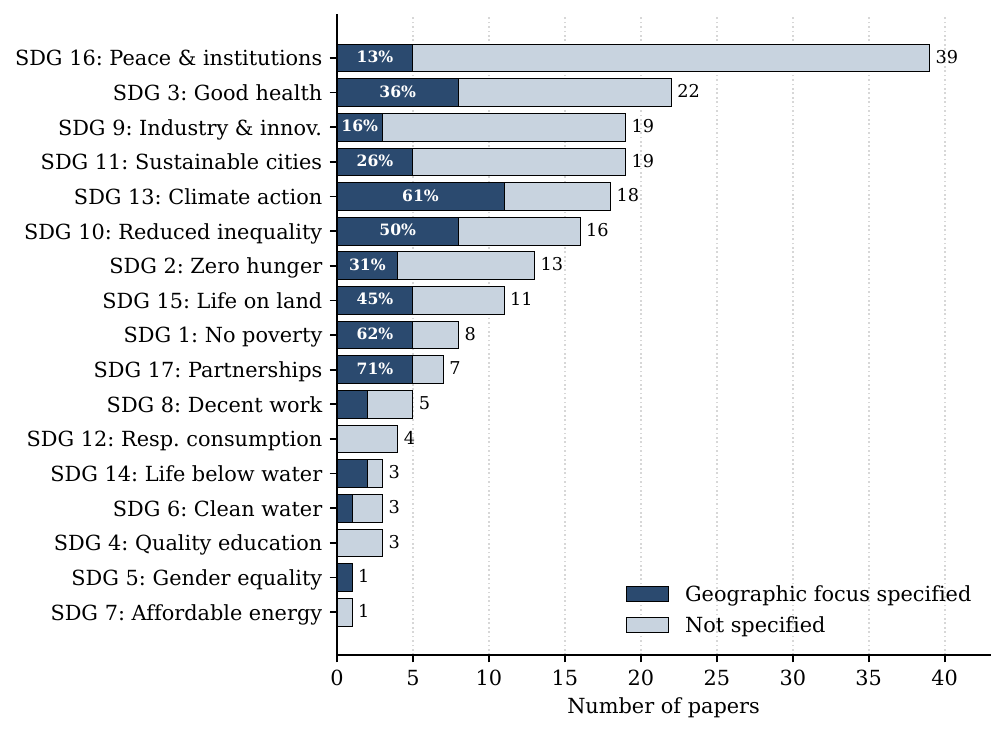}
    \caption{Distribution of corpus papers across Sustainable Development
    Goals (SDGs). The literature is concentrated on a subset of SDGs,
    especially SDG~16 and SDG~3, while several goals receive only sparse
    coverage.}
    \label{fig:sdg-distribution}
\end{figure*}

The SDG distribution reveals concentration around a subset of goals
(Figure~\ref{fig:sdg-distribution}). SDG~16 (peace, justice, and strong
institutions) appears most frequently with 39 papers, followed by SDG~3 (good
health, 22), SDGs~9 and 11 (industry/innovation and sustainable cities,
19--20 each), SDG~13 (climate action, 18), and SDG~10 (reduced inequality,
16). Several SDGs receive sparse coverage: SDGs~5 and 7 appear in 1 paper
each; SDGs~4, 6, and 14 in 3 each; SDG~12 in 4.

The point is not that every SDG should receive equal attention. Some goals
may be more naturally aligned with agentic methods; others require physical
infrastructure, field partnerships, or data that are difficult to obtain.
The problem is unexplained unevenness. Why do some goals become frequent
sites of agentic intervention while others remain peripheral? Is it because
agentic AI is genuinely unsuitable, because data and partners are
unavailable, because funding pulls elsewhere, or because some communities'
problems are less visible to AI researchers? These are empirical and
normative questions the literature rarely asks.

\subsection{Geographic Underspecification}
\label{sec:geo-underspec}

The clearest moral-geographic finding is the absence of geographic specificity.
Across the corpus, 82 of 112 papers (73\%) specify no geographic focus. Of the
remaining 30, 16 claim a broad or global scope, and only 14 name a concrete
country, city, region, or population---fewer than one in eight papers.

This is not a minor reporting issue. Social-good interventions are
context-dependent in ways that make geographic abstraction a substantive
problem, not a metadata gap. A disaster response agent depends on local
emergency infrastructure, communication networks, and institutional authority.
An agriculture agent depends on crop type, land tenure, and market access. A
governance agent depends on legal systems, administrative capacity, and rights
regimes. Without specifying where a system is meant to operate, it is
difficult to evaluate whether it is feasible, beneficial, or accountable.

\begin{figure*}[t]
    \centering
    \includegraphics[width=0.95\linewidth]{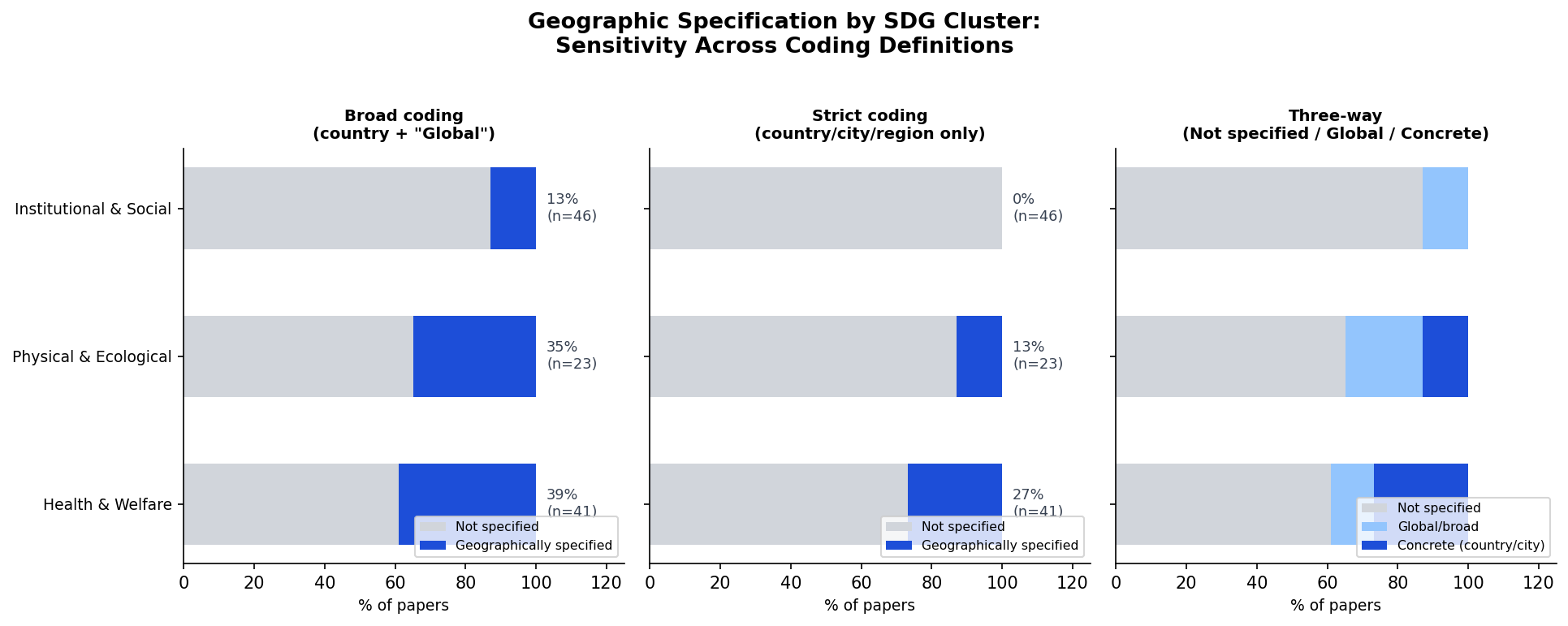}
    \caption{Sensitivity analysis for geographic specification coding.
    ``Global'' is not treated as geographic specificity, but as a distinct
    category of moral abstraction.}
    \label{fig:sensitivity-geo}
\end{figure*}

Crucially, claims of \emph{global} applicability do not resolve this
problem---they deepen it. A paper proposing an agentic system for ``global''
peace, governance, or institutional reform without naming a governance
setting, legal regime, or affected community is making a stronger
universalist claim, not a more cautious one. ``Global'' is therefore not
coded as geographic specification in our strict analysis; it is treated as
its own category of moral abstraction (see Figure~\ref{fig:sensitivity-geo}).

Geographic underspecification also obscures power: papers that describe
agents for ``sustainable cities'' or ``social policy'' without naming
relevant communities risk treating social good as an abstract category
rather than a situated relationship. Missing geography is not merely
missing metadata; it is a symptom of weak accountability.

\subsection{A Moral-Geographic Asymmetry Across SDGs}
\label{sec:asymmetry}

The 73\% marginal figure masks a sharper pattern. We grouped SDGs into three
clusters---\emph{health and welfare} (1--3), \emph{physical and ecological}
(6, 7, 11, 13--15), and \emph{institutional and social} (4, 5, 8--10, 12,
16, 17)---and assigned each paper to the cluster of its numerically lowest
SDG. Geographic specification is highly unevenly distributed across these
clusters (Table~\ref{tab:sensitivity}; Figure~\ref{fig:sensitivity-geo}).

\begin{table*}[t]
\centering
\small
\begin{tabular}{llccc ccc}
\toprule
& & \multicolumn{3}{c}{\textbf{Geographic specification rate}} &
  \multicolumn{3}{c}{\textbf{Test statistics}} \\
\cmidrule(lr){3-5}\cmidrule(lr){6-8}
\textbf{Clustering} & \textbf{Geo} &
\textbf{Inst.} & \textbf{Health} & \textbf{Phys/Eco} &
$\chi^2$(2) & $p$ & $V$ \\
\midrule
Original & Broad
  & 6/47 (13\%) & 14/38 (37\%) & 10/25 (40\%)
  & 8.78 & .012 & .283 \\
\textbf{Original} & \textbf{Strict}
  & \textbf{0/47 (0\%)} & \textbf{6/38 (16\%)} & \textbf{3/25 (12\%)}
  & \textbf{7.60} & \textbf{.022} & \textbf{.263} \\
\midrule
SDG~11 $\to$ inst. & Broad
  & 10/64 (16\%) & 14/38 (37\%) & 6/8 (75\%)
  & 15.32 & .000 & .373 \\
\textbf{SDG~11 $\to$ inst.} & \textbf{Strict}
  & \textbf{3/64 (5\%)} & \textbf{6/38 (16\%)} & \textbf{0/8 (0\%)}
  & \textbf{4.68} & \textbf{.096} & \textbf{.206} \\
\midrule
SDG~8 $\to$ health & Broad
  & 5/45 (11\%) & 15/40 (38\%) & 10/25 (40\%)
  & 10.08 & .006 & .303 \\
\textbf{SDG~8 $\to$ health} & \textbf{Strict}
  & \textbf{0/45 (0\%)} & \textbf{6/40 (15\%)} & \textbf{3/25 (12\%)}
  & \textbf{6.97} & \textbf{.031} & \textbf{.252} \\
\midrule
Single-SDG ($n=49$) & Broad
  & 1/30 (3\%) & 3/8 (38\%) & 3/11 (27\%)
  & 7.97 & .019 & .403 \\
\textbf{Single-SDG ($n=49$)} & \textbf{Strict}
  & \textbf{0/30 (0\%)} & \textbf{2/8 (25\%)} & \textbf{0/11 (0\%)}
  & \textbf{10.69} & \textbf{.005} & \textbf{.467} \\
\bottomrule
\multicolumn{8}{p{0.92\linewidth}}{\footnotesize
  \textit{Broad coding}: country, city, region, named population, or
  global/broad scope (e.g., ``Sub-Saharan Africa'').
  \textit{Strict coding}: country, city, region, or named population only;
  global claims excluded.
  Variant rows show alternative SDG cluster assignments. $V$ = Cram\'{e}r's $V$.
  $\dagger$~SDG~11~$\to$~inst.\ strict coding: $\chi^2$ unreliable with
  $n = 8$ in the physical/ecological cell after reassignment; the directional
  pattern holds but $p = .096$.}\\
\end{tabular}
\caption{Robustness of the moral-geographic asymmetry across geographic
coding schemes and cluster assignments. Seven of eight configurations
yield $p < .05$; the exception (SDG~11~$\to$~inst., strict coding) reflects
low cell counts ($n = 8$) after reassignment rather than a reversal of
the pattern. Under strict coding across all other configurations, no
institutional/social-policy paper names a concrete geographic context.
Bold rows indicate strict-coding results.}
\label{tab:sensitivity}
\end{table*}

Under the broad coding---which treats ``global'' and ``Sub-Saharan Africa''
as specified---the institutional cluster specifies a geography 13\% of the
time versus 37--40\% for the other two clusters: $\chi^2(2) = 8.78$,
$p = .012$, Cram\'{e}r's $V = .283$. Under the strict coding---which counts
only papers naming a concrete country, city, region, or identified population,
excluding global claims---the institutional cluster drops to \textbf{0 of 47
papers} (0\%), while health/welfare remains at 16\%: $\chi^2(2) = 7.60$,
$p = .022$, $V = .263$. This result is robust across seven of eight
sensitivity configurations we tested, including two alternative cluster
assignments and an analysis restricted to single-SDG papers
(Table~\ref{tab:sensitivity}; Figure~\ref{fig:sensitivity-geo}). The single
exception---Variant~A under strict coding---reflects low cell counts
($n = 8$ in the physical/ecological cluster after SDG~11 reassignment)
rather than a reversal of the pattern; the directional asymmetry holds
in all eight configurations.

\begin{figure*}[t]
    \centering
    \includegraphics[width=0.7\textwidth]{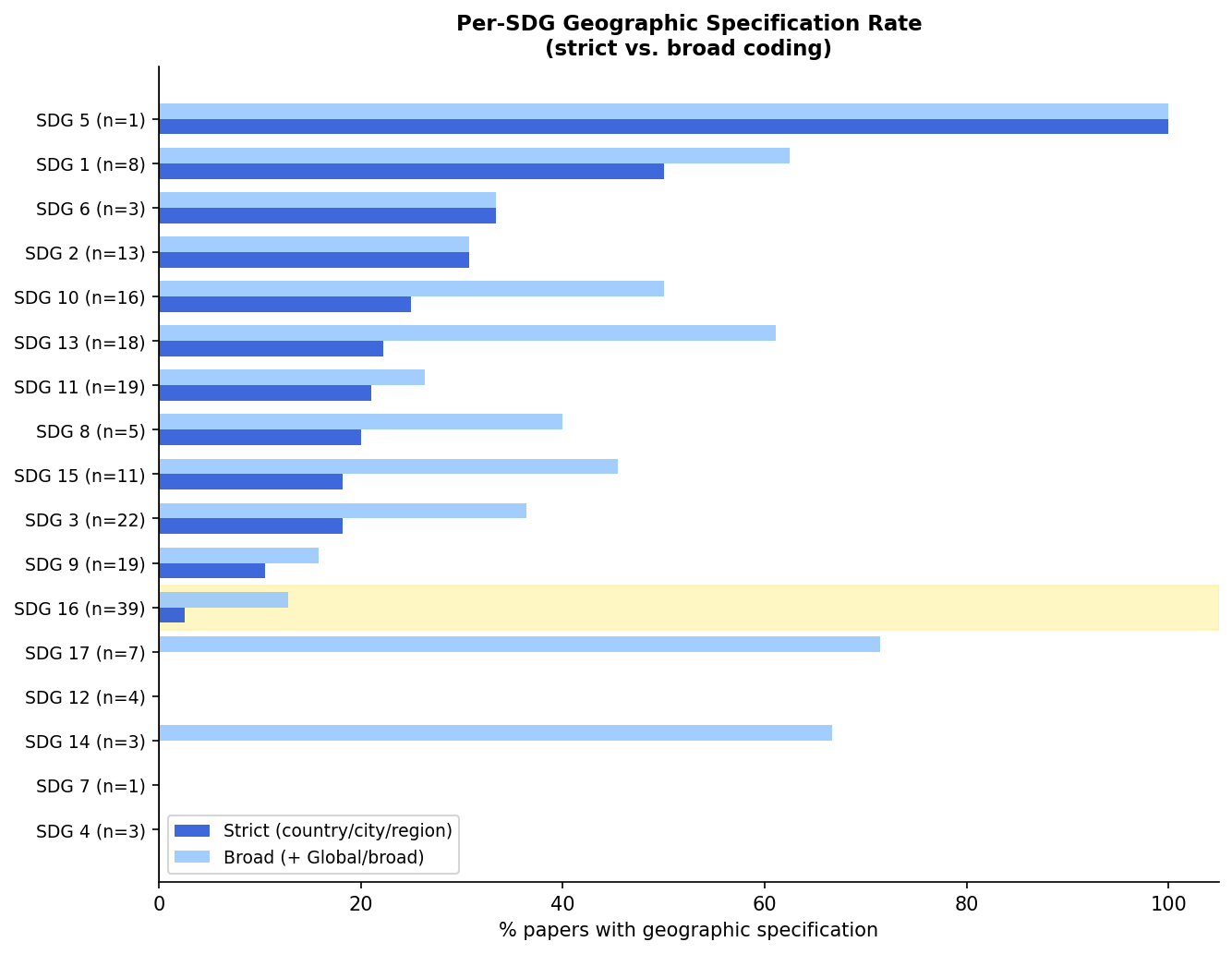}
    \caption{Per-SDG geographic specification rates under broad and strict
    coding. The institutional and governance-oriented SDGs, especially
    SDG~16, show substantially lower geographic specification than climate,
    poverty, and ecological domains.}
    \label{fig:persdg-sensitivity}
\end{figure*}

The per-SDG view sharpens this further
(Figure~\ref{fig:persdg-sensitivity}). SDG~16---the most-covered goal in
the corpus---has the lowest geographic-specification rate: 5 of 39 papers
(13\%) under broad coding, and just 1 of 39 (2.6\%) under strict coding.
SDG~9 follows at 16\% and 0\% respectively. By contrast, SDG~13 (climate
action) reaches 61\% under broad coding; SDG~1 (no poverty) reaches 62\%.

We interpret this as a \emph{moral-geographic asymmetry}: agentic AI for
social good treats institutional and social-policy domains as universal while
treating health, climate, and ecological domains as located. A paper
proposing an agentic system for peace, governance, or institutional reform is
roughly three times less likely under broad coding---and categorically less
likely under strict coding---to be tethered to a specific geography than a
paper proposing an agentic system for climate or food security. Yet
institutional interventions are precisely those whose feasibility,
legitimacy, and risk depend most sharply on local political, legal, and
cultural context.

This asymmetry is the central empirical contribution of this survey. It is
distinct from the simpler claim that AI4SG underspecifies context: the
pattern is directional, concentrated in the goals the literature most
frequently invokes as legitimating vocabulary, and strengthens rather than
weakens under stricter geographic definitions.

\section{The Distance Between Proposal and Practice}
\label{sec:distance}

\subsection{Deployment Status}

The corpus is dominated by conceptual and simulation-only work
(Figure~\ref{fig:evaluation-type}). Forty-six of the 112 papers are coded
as \emph{No (conceptual)} and 33 as \emph{Simulation only}. Only 28 papers
(25\%) report any real-world deployment or small-scale test: 23 small-scale
tests and 5 field deployments. Four papers are coded \emph{Various}.

Conceptual and simulation work is not without value. Early-stage systems
require framing and controlled evaluation before deployment, and in
high-stakes social domains premature deployment can cause harm. The concern
is not that every paper should deploy a system. The concern is that the
literature routinely invokes real-world social benefit while remaining at
the level of possibility or prototype. A simulation can test coordination
dynamics but cannot establish community benefit. An offline benchmark can
test task performance but cannot establish institutional feasibility. A
conceptual architecture cannot show it is appropriate for a particular
social context.

One clarification on the 25\% figure: the \emph{small-scale test} category
spans a range of evidentiary postures, from recruited-participant user
studies to small operational pilots. Under a stricter definition counting
only field deployments, the rate falls to 5 of 112 papers (4.5\%). Both
figures are informative; we report the broader count to remain conservative,
but the narrower figure better represents evidence of real-world impact.

\subsection{Evaluation Type}

\begin{figure}[t]
    \centering
    \includegraphics[width=0.95\linewidth]{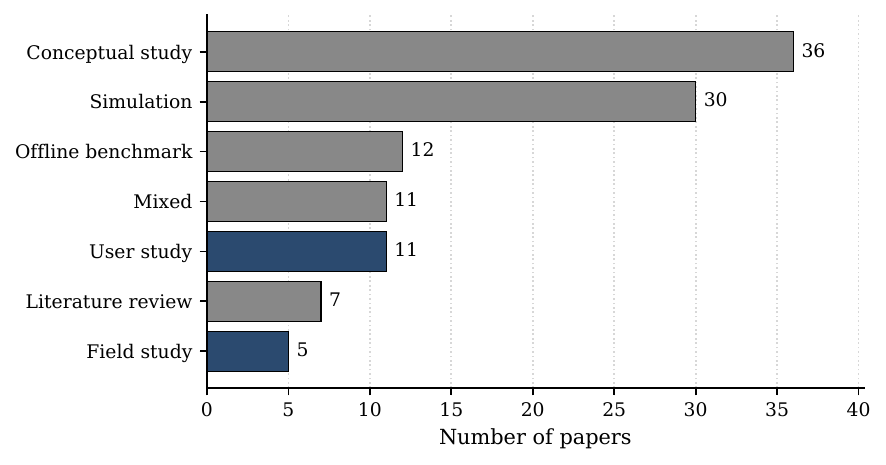}
    \caption{Distribution of evaluation types in the corpus. Most papers are conceptual studies or simulations, while user studies and field studies remain relatively rare.}
    \label{fig:evaluation-type}
\end{figure}

Evaluation practices mirror the deployment gap
(Figure~\ref{fig:evaluation-type}). The most common type is
\emph{Conceptual study} (36 papers), followed by \emph{Simulation} (30) and
\emph{Offline benchmark} (12). Only 11 papers report a \emph{User study},
and just 5 report a \emph{Field study}.

This matters because agentic systems are interactional. Their performance
depends not only on model outputs but on human trust, workflow integration,
institutional constraints, failure recovery, and contestability---properties
that static benchmarks cannot capture. A planning agent that produces
plausible plans may still fail when used by overburdened public servants. A
mental-health engagement agent may score well in controlled evaluation but
behave differently with vulnerable users over time. A disaster coordination
agent may work in simulation but fail when communication infrastructure
collapses.

The relevant question is not merely whether systems
work, but whether the evaluation design is adequate to the social claim.
A benchmark result supports a claim about task performance; it should not
stand as evidence of community benefit, institutional readiness, or
responsible deployment. Papers proposing consequential agentic roles should
either provide evidence appropriate to those roles or explicitly bound
their claims.

\subsection{The Deployment Gap is Field-Wide}
\label{sec:nulls}

We pre-specified three hypothesis tests connecting deployment status
(deployed-or-tested vs.\ conceptual-or-simulation) to the splits most
likely to concentrate the gap. None returned a significant effect.

\begin{itemize}
  \item \textbf{Geography $\times$ Deployment.} Papers naming a geography
    deploy or test at 8/28 (29\%) vs.\ 20/80 (25\%) for those that do not.
    Fisher's exact $p = .80$, OR = 1.20, $V = .01$.
  \item \textbf{Year $\times$ Deployment.} Post-2023 papers: 19/76 (25\%);
    pre-2023: 9/32 (28\%). Fisher's exact $p = .81$, OR = 1.17, $V = .01$.
  \item \textbf{Agent class $\times$ Deployment.} LLM-based architectures:
    14/54 (26\%); classical architectures: 10/46 (22\%); human--AI teams:
    3/7 (43\%). $\chi^2(2) = 1.46$, $p = .48$, $V = .12$.
\end{itemize}

The proposal-to-practice gap is not concentrated in any sub-community we
could identify; it is a \emph{field-wide condition}. Papers that name a
geography do not deploy at higher rates. The LLM-agent wave is neither more
conceptual nor more field-tested than the multi-agent-systems work it builds
on. The gap is structural, and this is a sharper critique than a localized
one would be: whatever the field claims, it has not established what its
systems do in practice across any cut of the data.

\subsection{Architecture and Accountability}

The corpus spans multi-agent systems (30 papers), agentic AI systems (23),
LLM agents (20), multi-agent LLMs (14), simulation agents (10), human--AI
teams (7), single-agent systems (6), and embodied agents (1)
(Figure~\ref{fig:evaluation-type}). Autonomy level is harder to characterize:
as noted in the \textit{Dual-LLM Annotation with Human Validation} section, cross-LLM agreement is only
$\kappa = 0.56$ (unweighted), indicating genuine ambiguity in how papers
describe the human--machine authority boundary. We treat the autonomy
distribution as descriptive and return to its implications in
the \textit{Accountability Gaps} section.

Architecture is often treated as a technical design choice; in social-good
settings it is also an accountability choice. Complex multi-agent LLM
systems may introduce more failure points, more opaque coordination
dynamics, and higher barriers to auditing than simpler advisory systems.
They may also require computational resources and monitoring capacity that
only well-resourced organizations can sustain. Simpler decision-support
systems may be less expressive but easier to audit, maintain, and integrate
into existing institutional workflows.

Architectural complexity is not inherently irresponsible, but in social-good
settings it should be justified: why is this level of autonomy or
coordination necessary, who can operate and maintain the system, how are
failures detected, and how can affected communities contest its outputs?
The $\kappa = 0.56$ autonomy-ambiguity finding from the \textit{Dual-LLM Annotation with Human Validation} section
suggests the field can be considerably more precise about where
system-initiated action ends and human-authorized action begins.

Together, Sections \textit{The Moral Geography of Agentic AI for Social Good} and \textit{The Distance Between Proposal and Practice}
describe not simply a deployment gap but an \emph{evidentiary gap}. The
literature increasingly proposes agentic systems for planning, coordination,
persuasion, monitoring, and resource allocation, yet most remain conceptual
or simulated, evaluated without the evidence needed to substantiate their
social-good claims---and where contextual specification is supplied, it is
supplied unevenly, concentrated in physical and ecological domains while
abstracting away from the institutional ones where contextual stakes are
highest.

\section{Accountability Gaps}
\label{sec:gaps}

Across the corpus, we identify five accountability gaps that limit the
maturity of agentic AI for social good. These are field-level patterns,
not indictments of individual papers.

\subsection{Under-Attended Goals}
\label{sec:gap-goals}

The corpus concentrates attention around a subset of SDGs while others
receive sparse coverage (\textit{The Moral Geography of Agentic AI for Social Good} section). Uneven distribution
is not inherently wrong---some goals may align more naturally with agentic
methods, others require data or partnerships that are genuinely difficult
to obtain. The problem is \emph{unexplained} unevenness. A field
presenting itself as a general-purpose technology for social good should
account for why certain problems become frequent sites of agentic
intervention while others remain peripheral. Each possible
explanation---missing data, funding gaps, reduced visibility of certain
communities' problems---carries different ethical implications.

\subsection{Missing Geographic Specificity, Unevenly Distributed}
\label{sec:gap-geo}

Most papers do not name a geographic focus, making it difficult to assess
institutional feasibility, affected communities, or local accountability.
Geography is not incidental metadata: it determines infrastructure,
language, regulation, and institutional capacity on which an intervention
depends.

The pattern in the \textit{A Moral-Geographic Asymmetry Across SDGs} section sharpens this into a pointed
diagnostic. Under strict coding, the rate for institutional papers drops
to \textbf{0\%}: no paper proposing an agentic system for peace, justice,
or governance names a concrete country, city, or population---precisely
backwards relative to where ethical risks are highest. Claims of global
applicability do not resolve this: a system described as ``global''
without naming a governance setting or affected community is making a
stronger universalist claim, not a more cautious one.

\subsection{Limited Field and User Evaluation}
\label{sec:gap-evaluation}

Only 28 of 112 papers (25\%) report any real-world deployment or
small-scale test, and only 5 conduct a field study (section \textit{The Distance Between Proposal and Practice});
under a stricter definition, the rate falls to 4.5\%. Agentic systems are
interactional---how humans respond to an agent, how institutions
incorporate it, and how failures propagate cannot be captured by static
benchmarks or one-time simulations. The null findings (section \textit{The Distance Between Proposal and Practice}) make this gap harder to dismiss: the deployment
shortfall is uniform across geography, year, and agent architecture. It is
a field-wide condition.

\subsection{Opaque Architectural and Autonomy Choices}
\label{sec:gap-architecture}

Architecture is often treated as a technical default in social-good
settings, but it is also an accountability choice: it determines who can
deploy, audit, and contest a system. Multi-agent LLM systems may require
computational resources unavailable to community organizations; simpler
advisory systems may be easier to audit and integrate into existing
workflows. A related issue is how autonomy is \emph{described}:
cross-LLM agreement on \texttt{Autonomy\_Level} reached only
$\kappa = 0.56$, the lowest of any field,
reflecting genuine ambiguity in how papers specify the boundary between
system-initiated and human-authorized action. Both architectural
complexity and autonomy level should be justified explicitly---not as
technical defaults, but as accountable choices. The default should not be
maximum agency, but \emph{accountable} agency.

\subsection{Weak Failure Documentation}
\label{sec:gap-failure}

The corpus contains little systematic discussion of failure modes,
non-deployments, or harms discovered during testing. A field aiming to
support high-stakes social outcomes cannot learn when agentic AI is
inappropriate or unwanted if it reports primarily successful prototypes.
Failure documentation should be treated as a contribution---a failed
deployment can reveal missing infrastructure or community mistrust
invisible in simulation. Closing this gap requires venue-level support:
reviewers should actively reward careful accounts of failure.

\medskip
\noindent These five gaps are connected: under-attended goals require
community partnerships, which depend on naming a place; missing geography
makes evaluation less meaningful; limited field evaluation obscures
successes and failures alike; opaque architecture makes accountability
harder to assign; and weak failure documentation prevents collective
learning. The challenge is to move from agentic AI as a \emph{language
of possibility} to a \emph{practice of accountable sociotechnical
intervention}.

\section{Recommendations}
\label{sec:recommendations}

The five gaps identified above motivate five practices and a minimal
reporting standard.

\subsection{Specify the Target Context}
Papers should identify the communities, institutions, or deployment
settings the system is intended for. Context shapes problem definition,
institutional constraints, failure risks, and success criteria
\citep{selbst2019fairness,mohamed2020decolonial,birhane2021algorithmic}.
If a system is intended to be general, authors should specify what kind
of generality is claimed---across tasks, languages, institutions, or
resource settings. This recommendation is sharpest for governance and
institutions---precisely where our analysis finds zero geographic
specification under strict coding.

\subsection{Match Evidence to the Strength of the Claim}

Evaluation should match the social claim being made. Claims of real-world
benefit require more than technical evidence: user studies, field studies,
or longitudinal deployment evidence. When stronger evidence is absent,
authors should bound their claims---simulations support coordination
dynamics, not community benefit; benchmarks support task performance, not
institutional readiness \citep{green2019good,tomasev2020ai}.

\subsection{Engage Affected Stakeholders Early}

Stakeholders should be engaged before core design choices are
fixed---not after architecture and evaluation metrics have been selected.
They should help define the problem, set success criteria, and shape
oversight mechanisms \citep{bondi2019roles,birhane2022power}. Affected
people should have a role in deciding whether a system should exist and
how it is governed---not only as users in a final evaluation.

\subsection{Justify Autonomy and Architecture}

Authors should justify why the system needs to be agentic, why it
requires the proposed autonomy level, and why its architecture fits the
target setting. The default should not be maximum agency but
\emph{accountable} agency: papers should specify what is delegated to
the agent, how failures are detected, and how responsibility is assigned
when harm occurs \citep{floridi2018ai4people,raji2020closing,dobbe2021system}.
The $\kappa = 0.56$ autonomy-ambiguity finding suggests the field can be
considerably more precise about where system-initiated action ends and
human-authorized action begins.

\subsection{Document Failures and Non-Deployments}

The field needs systematic documentation of failures, non-deployments,
and unsuccessful simulation-to-practice transfers. A failed deployment
may reveal missing infrastructure or community mistrust invisible in
simulation; a non-deployment decision can demonstrate responsible
restraint. Venues and reviewers should actively reward such accounts.

\subsection{A Minimal Reporting Standard}
\label{sec:reporting-standard}

Papers on agentic AI for social good should report: (1)target
context and affected community; (2) social claim and supporting
evidence; (3) stakeholder role in problem definition and
oversight; (4) autonomy and architecture justification,
including failure detection and contestability; and
(5) limitations and failure modes. This standard is
intentionally modest---it requires authors to state what kind of
social-good claim they are making and what evidence, context, and
accountability mechanisms support it. Adopted broadly, it would shift
the field from asking only what agents \emph{can} do to asking what
agentic AI for social good \emph{owes} to the people and institutions
in whose name it is built.

\section{Conclusion}
\label{sec:conclusion}
Agentic AI for social good is expanding rapidly, but expansion is not
maturity. Our survey of 112 papers shows the field is currently weak on
context, evidence, and accountability: most papers do not name a geographic
setting, only one in four reports any real-world deployment, and the
deployment gap is uniform across architecture, era, and context.

The sharpest pattern is not the breadth of these absences but their shape.
Contextual underspecification is concentrated in the institutional and
social-policy SDGs the literature most invokes as legitimating
vocabulary---peace, governance, inequality---precisely where the literature
is least willing to name a place. This moral-geographic asymmetry is the
central contribution of this survey, distinct from the simpler claim that
AI4SG underspecifies context.

The path forward is not to deploy more agents but to build more accountable
ones: systems whose communities are named, whose autonomy is justified,
whose evaluations match their claims, and whose failures are documented.
Agentic AI for social good should not merely ask what agents can do. It
should ask what it owes to the people and places in whose name it is
conducted.

\section*{Limitations}

\label{sec:limitations}

This survey has several limitations.

First, the corpus may not include all relevant work on agentic AI for
social good. Although we followed a PRISMA-style search procedure across
four academic discovery platforms with SDG-organized queries, the field is
terminologically diffuse. Some relevant papers may use domain-specific
language without describing their systems as agents, agentic AI, or
multi-agent systems. Conversely, some papers may use agentic language for
systems whose degree of agency is limited. Our both-LLMs-agree filter is conservative about exclusion: a paper is
dropped as non-agent only when both annotators independently flag it as
such, supplemented by human eligibility review. This rule retains some
borderline cases that a stricter filter would have removed, which we read
as the appropriate direction of error for a field-level survey.

Second, SDG coding is interpretive. Some papers align with multiple goals,
while others address SDG-relevant domains without explicitly citing the
SDG framework. We therefore treat SDG labels as an analytical lens rather
than as definitive claims about author intent. The purpose of the SDG
analysis is not to assign papers to a single correct goal, but to examine
how attention is distributed across a widely recognized vocabulary of
social benefit. The multi-label coding produces some attribution
ambiguity: when a paper aligns with both an institutional and a
health/ecological SDG, our cluster assignment uses the first matching SDG
in the paper's primary label set. Alternative tie-breaking rules would
shift individual papers between clusters but not the cluster-level
pattern.

Third, our coding relies on how papers present themselves in titles and
abstracts. This is appropriate for analyzing field-level framing, but it
may miss details available only in full texts, appendices, code
repositories, or deployment reports. In particular, geographic focus is
coded based on what the paper specifies, not what the authors may have
intended. A paper coded as \emph{Not specified} may still have been
motivated by a concrete setting, but if that setting is not named in the
title or abstract, it is not visible as part of the paper's public
framing.

Fourth, deployment status and evaluation type simplify a more complex
continuum of evidence. A small-scale test, field study, simulation, and
benchmark can each vary substantially in rigor, duration, stakeholder
involvement, and relevance to real-world use. Our labels should therefore
be read as high-level indicators of evidentiary posture rather than
complete evaluations of methodological quality.

Fifth, the analysis does not directly measure social impact. We do not
evaluate whether deployed systems improved outcomes, whether communities
benefited, or whether harms occurred after deployment. We analyze how the
literature frames, evaluates, and contextualizes its proposed systems.
Our claims concern the distribution of attention, context, and evidence
in the research corpus, not the actual downstream impact of every system
reviewed.

Sixth, our dual-LLM consensus procedure adjudicates LLM disagreements
only within the human-validated 50-paper subsample. For the remaining 62
papers in the consensus corpus, we adopt the field-appropriate LLM
without cell-by-cell adjudication. This is a deliberate budget tradeoff:
manual adjudication of all $\sim$400 cell-level disagreements was
infeasible, and using a single LLM throughout would have lost the
human-validation grounding entirely. We report cross-LLM agreement
statistics (Table~\ref{tab:iaa-llm}) and human--LLM agreement
(Table~\ref{tab:iaa-human}) to characterize the reliability of the
LLM-only portion of the corpus.
Because Geographic Focus is central to our main analysis, we additionally inspected all papers contributing to the specified/not-specified split for label validity after consensus construction.
The binary geographic-focus label used in the main analysis was manually checked for all 112 included papers after consensus construction.

Seventh, \texttt{Autonomy\_Level} is the noisiest field in our coding,
with unweighted $\kappa = 0.56$ between the two LLM annotators. We treat
this as a substantive finding about the field rather than a failure of
the annotation procedure: autonomy
claims in titles and abstracts are themselves ambiguous. We therefore
read autonomy distributions descriptively, do not draw inferential
conclusions from them, and use only the more reliable
\texttt{Real\_World\_Deployment} and
\texttt{Geographic\_Focus} fields for our
main hypothesis tests.

These limitations shape the scope of our argument. We do not claim that
the field's distribution of attention is wrong, that any specific SDG
deserves more research, or that any specific agentic system should or
should not be deployed. We claim that the field's distribution of
attention, context, and evidence is ethically relevant and currently
underexamined, and that one specific pattern within it---the
moral-geographic asymmetry between institutional and physical SDGs---is
sharp enough to merit explicit attention. The purpose of this survey is
not to close the debate about agentic AI for social good, but to make its
assumptions, absences, and evidentiary standards more visible.

\section*{Data and Code Availability}
The consensus corpus, annotation codebook, and analysis scripts are available at:
\url{https://github.com/PoliNemkova/agentic_social_good}.

\bibliography{aaai2026}

\appendix

\clearpage
\section{Appendix 1. Corpus Construction and Screening Protocol}
\label{app:corpus-construction}

This appendix provides additional detail on the corpus-construction procedure summarized in Section \textit{Method}. We followed a PRISMA-style systematic review procedure to identify papers on agentic AI for social good. The goal was not to exhaust every possible paper using agent-related terminology, but to construct a transparent, SDG-organized corpus suitable for analyzing how the literature frames social-good claims, deployment status, and geographic context.

\subsection{Search Sources}

We searched four academic discovery platforms: Google Scholar, ACL Anthology, Semantic Scholar, and Elicit. These sources were selected to capture work across NLP, AI, multi-agent systems, human--AI collaboration, and interdisciplinary AI-for-social-good research. Because agentic AI for social good is terminologically diffuse, we did not rely on a single keyword such as ``agent.'' Instead, we organized the search around the 17 United Nations Sustainable Development Goals (SDGs), using both application-oriented and task-oriented queries for each goal.

\subsection{Search Strategy}

For each SDG, we constructed search queries combining social-good domains with agentic-system terminology. Application-oriented queries targeted concrete domains such as clinical triage agents, disaster response agents, agricultural advisory agents, climate adaptation agents, educational tutoring agents, misinformation verification agents, and public-service decision-support agents. Task-oriented queries targeted agentic functions such as planning, coordination, monitoring, resource allocation, forecasting, verification, persuasion, retrieval, and human--AI collaboration.

This dual strategy was intended to retrieve papers that explicitly present themselves as agentic systems, as well as papers embedded in domain-specific literatures where agent architectures may not be foregrounded in the title. Retrieved records were merged and deduplicated before screening. The search yielded 177 candidate papers.

\subsection{Inclusion Criteria}

Papers were included if they satisfied all three criteria below:

\begin{enumerate}
    \item \textbf{Agentic system criterion:} The paper described, analyzed, evaluated, or surveyed an AI agent, multi-agent system, LLM agent, multi-agent LLM system, simulation agent, embodied agent, agentic AI workflow, or human--AI collaboration involving goal-directed behavior beyond static prediction or classification.
    
    \item \textbf{Social-good criterion:} The paper targeted a domain plausibly connected to social good, sustainable development, or one or more SDGs, including but not limited to public health, education, climate, agriculture, disaster response, humanitarian decision-making, governance, online safety, social policy, sustainability, or urban planning.
    
    \item \textbf{Codability criterion:} The title and abstract provided enough information to code the paper's domain, agent type, and social-good orientation.
\end{enumerate}

We define an agentic AI system broadly as a system that performs or supports goal-directed behavior over time, including systems that plan, coordinate, monitor, verify, persuade, allocate resources, forecast, retrieve information, use tools, or support human decision-making.

\subsection{Exclusion Criteria}

Papers were excluded if they met any of the following criteria:

\begin{enumerate}
    \item The paper was policy-only or conceptual with no identifiable agentic system.
    \item The paper described a non-agentic machine-learning system, such as a static classifier, predictor, recommender, or content generator without an agentic role.
    \item The system was commercially motivated without a clear social-good objective.
    \item The paper was purely theoretical without an identifiable system, application, or evaluation target.
    \item The title and abstract were too generic to establish relevance to agentic AI for social good.
    \item The paper fell outside the 2015--2026 date range.
\end{enumerate}

After dual-LLM screening and human eligibility review, 65 papers were excluded: 55 by both-LLM consensus as non-agent, 8 during human eligibility review, and 2 because they were published before 2015. The final corpus contains 112 papers.

\subsection{Screening and Eligibility Review}

Each candidate paper was screened using title and abstract information. This design choice reflects the paper's focus on public-facing field-level framing: we analyze how papers present their social-good claims, target contexts, and agentic contributions in the metadata most visible to readers, funders, and downstream researchers. We do not claim that title and abstract information exhausts all system details available in full papers; rather, we treat title and abstract framing as evidence of what the field makes visible when presenting agentic AI for social good.

Eligibility was first assessed using the dual-LLM procedure described in Section \textit{Method}. Papers that both LLM annotators marked as non-agentic or irrelevant were excluded unless flagged for human review. Borderline cases were retained when the abstract plausibly described goal-directed agentic behavior in a social-good domain. This conservative inclusion rule was chosen to avoid prematurely excluding emerging or hybrid forms of agentic AI whose terminology may not yet be standardized.

\subsection{PRISMA-Style Summary}

The corpus-construction process can be summarized as follows:

\begin{itemize}
    \item \textbf{Identification:} Searches across Google Scholar, ACL Anthology, Semantic Scholar, and Elicit using SDG-organized application-oriented and task-oriented queries.
    \item \textbf{Candidate records:} 177 papers after merging and deduplication.
    \item \textbf{Screening and eligibility review:} Candidate papers screened for agentic-system relevance, social-good orientation, and codability from title and abstract.
    \item \textbf{Excluded records:} 65 papers excluded: 55 by both-LLM consensus as non-agent, 8 during human eligibility review, and 2 pre-2015 papers.
    \item \textbf{Final corpus:} 112 papers included in the structured survey.
\end{itemize}

\subsection{Scope of Claims}

The resulting corpus should be read as a structured sample of how agentic AI for social good is publicly framed in academic literature, not as a complete census of all deployed systems or all domain-specific agentic research. Some relevant systems may be described in technical reports, NGO documentation, government pilots, software repositories, or full-text sections not visible from title and abstract. Conversely, some included papers may use agentic language for systems whose degree of autonomy is limited. These limitations motivate our conservative interpretation: the analysis concerns the distribution of attention, context, evidence, and visible accountability claims in the academic literature, rather than the full downstream impact of every system reviewed.


\end{document}